\documentclass[onecollarge]{svjour2}
\smartqed  
\usepackage{graphicx}
\usepackage{xcolor}

\journalname{Few-Body Systems (EFB22)}
\begin{document}

\title{Quark mass functions and pion structure in Minkowski space 
}


\author{Elmar P. Biernat \and Franz Gross \and M.~T.~Pe\~na \and Alfred Stadler        
}


\institute{Elmar P. Biernat and M.~T.~Pe\~na  \at
             Centro de F\'isica Te\'orica de Part\'iculas (CFTP), 
Instituto Superior T\'ecnico (IST), Universidade de Lisboa\\ 
Av. Rovisco Pais, 1049-001 Lisboa, Portugal \\
                            \email{elmar.biernat@tecnico.ulisboa.pt}     
                            \and
         Franz Gross \at Thomas Jefferson National Accelerator Facility (JLab), Newport News, VA 23606, USA
         \and
        Alfred Stadler \at Departamento de F\'isica da Universidade de \'Evora, 7000-671 \'Evora, Portugal and \\ Centro de F\'isica Nuclear da Universidade de Lisboa (CFNUL), 1649-003 Lisboa, Portugal}
\date{Received: date / Accepted: date}

\maketitle

\phantom{0}

\begin{abstract}We present a study of the dressed quark mass function and the pion structure in Minkowski space using the Covariant Spectator Theory (CST). The quark propagators are dressed with the same kernel that describes the interaction between different quarks.
We use an interaction kernel in momentum space that is a relativistic generalization of the linear confining $q\bar q$ potential and a constant potential shift that defines the energy scale. The confining interaction has a Lorentz scalar part that is not chirally invariant by itself but decouples from the equations in the chiral limit and therefore allows the Nambu--Jona-Lasinio (NJL) mechanism to work. We adjust the parameters of our quark mass function calculated in Minkowski-space to agree with LQCD data obtained in Euclidean space. Results of a calculation of the pion electromagnetic form factor in the relativistic impulse approximation using the same mass function are presented and compared with experimental data. 

\keywords{Quark mass function\and Pion form factor \and Covariant Spectator Theory \and Confinement \and Spontanteous chiral symmetry breaking}
\end{abstract}

\section{Introduction}
\label{intro}
To understand the connection between the observed properties of mesons and baryons and the underlying QCD quark-gluon dynamics is one of the greatest challenges in Hadron Physics. At present, models are needed to establish a link between the results from lattice QCD (LQCD) simulations \cite{EdwardsGuo} and the existing experimental data, as well as the ones from future experimental programs, for instance at JLab and FAIR. Any realistic model of hadrons must incorporate two key features that dominate the real world: (i) Dynamical chiral-symmetry breaking -- responsible for the spontaneous generation of a constituent quark mass, which, in the chiral limit, is linked to the existence of the zero-mass pion through the famous NJL mechanism, and (ii) quark confinement. 

Our model differs in two important aspects from the very successful Dyson-Schwinger approaches (see for example \cite{Qin2011} and references therein): (i) we work in a framework formulated in \textit{physical} Minkowski space instead of \textit{unphysical} Euclidean space, which allows, for instance, the straightforward calculation of transition form factors in the timelike region, and (ii) we are able to accomodate a Lorentz-scalar confining interaction as suggested by phenomenological approaches and lattice calculations~\cite{scalarconf} (although those indications are not definitive). Such a scalar confining interaction can be included without spoiling chiral symmetry, as it decouples in the chiral limit, allowing the NJL mechanism to work. 

A more detailed description of our work, which can be viewed as an extention, improvement and application of earlier work by Gross, Milana and \c{S}avkli \cite{GrossMilanaSavkli} based on the Covariant Spectator Theory (CST)~\cite{CST}, can be found in Refs.~\cite{Biernat:2013fka,Biernat:2013aka}.

\section{Quark mass function}
\label{sec:1}
In Ref.~\cite{Biernat:2013fka} we propose a manifestly covariant model for the $q\bar q$ interaction that incorporates both spontaneous chiral-symmetry breaking and confinement. Our model is particularly suited for the treatment of the light mesons such as the pion. In this case a four-channel CST formulation is employed which is invariant under charge conjugation. On the other hand, in the non-relativistic limit, the CST equation reduces to the Schr\"{o}dinger equation. This makes our approach suitable for a unified description of all $q\bar q$ mesons. 

We use a kernel consisting of the sum of a momentum-space $\delta$-function with a pure vector Lorentz structure and a mixed scalar-vector confining interaction. Our kernel includes (strong) phenomenological form factors $h(p^2)$~\cite{Gro87,Gro92,GrossMilanaSavkli,Gro96} for each quark line associated with momentum $p$. They can be regarded an effective description of the infinite sum of overlapping gluon loop corrections to the interaction vertices and they also provide convergence in loop integrals. 

Spontaneous chiral-symmetry breaking is included via a NJL-type mechanism: A self-consistency condition ensures that the CST equation for a pseudoscalar bound state has a zero-mass solution (Goldstone pion) in the chiral limit of a vanishing current quark mass $m_0$. A finite dressed quark mass is then generated dynamically through the self-interactions of the quark with the $q\bar q$ interaction kernel. In previous models~\cite{GrossMilanaSavkli} the quark mass function has been approximated by a phenomenological function. In the present approach we actually calculate the mass function directly from the $q\bar q$ interaction kernel, which makes our model completely self-consistent. 

\begin{figure}[h!] \begin{center} 
    \includegraphics[height=6cm]{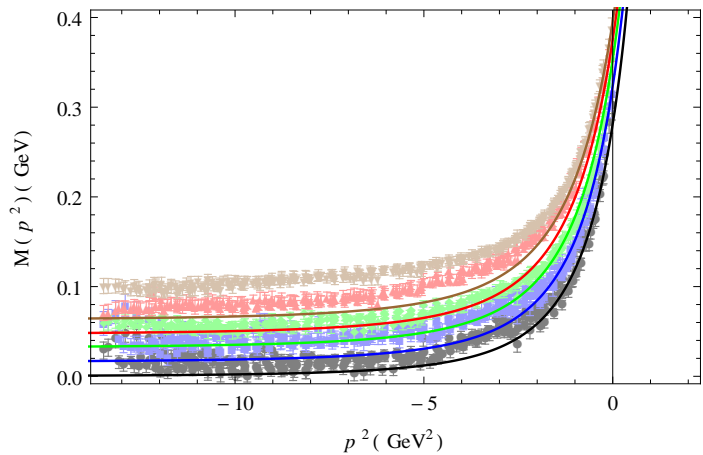} 
\caption{The quark mass functions compared with LQCD data~\cite{Bowman:2005vx} for different $m_0$. The five curves and (extrapolated) data sets from bottom to top correspond to bare quark masses $m_0=0$ (black, blobs),  $m_0=0.016$ GeV (blue, squares), $m_0=0.032$ GeV (green, diamonds), and $m_0=0.047$ GeV (red, triangles) and $m_0=0.063$ GeV (brown, inverted triangles).}\label{fig:1}\end{center}
\end{figure}

For a particular Lorentz scalar-vector mixing the confining potential does not contribute to the dynamical quark mass generation. In this case our mass function involves only 3 free parameters, the constituent quark mass $m_\chi$ in the chiral limit, a mass parameter $M_g$ from the quark form factors and the strength $C$ of the $\delta$-function kernel. The parameters $m_\chi$ and $M_g$ are fixed in the chiral limit by a fit to the lattice data~\cite{Bowman:2005vx} extrapolated to $m_0=0$, which gives $m_\chi=0.308~\mathrm{GeV}$ and $M_g=1.734~\mathrm{GeV}$. The mass function for different values of $m_0$ is then found by solving the corresponding gap equation, with $C$ as a function of $m_0$ chosen appropriately to fit the data. The mass function result reads~\cite{Biernat:2013aka} 
\begin{eqnarray} 
 M(p^2)= \left(m_\chi+12 m_0\right)\,h^2(m^2)h^2(p^2)+m_0\,,\label{eq:1}
\end{eqnarray}
where
\begin{eqnarray}
 h(p^2)=\left(\frac{\Lambda_\chi^2-m_\chi^2}{\Lambda^2-p^2}\right)^2\, \label{eq:2}
\end{eqnarray}
are the (strong) quark form factors with $\Lambda=m+M_g$ and $\Lambda_\chi=m_\chi+M_g$. The lattice data (calculated in Euclidean space) are compared with our Minkowski-space results at negative $p^2$, as shown in Figure~\ref{fig:1}.

When fixing our mass function parameters we have fitted only the first 50 data points (in the low-$p^2$ region between 0 and $-1.94~\mathrm{GeV}^2$) of the chiral-limit extrapolation of the lattice data. The reason for this are finite lattice spacing effects, responsible for the deviation of the lattice data from the correct asymptotic UV-behavior, that are expected to be small in this low-$p^2$ region. Figure~\ref{fig:1} shows, that for small $m_0$ and for small $p^2$ our mass function provides a good fit to the lattice data.

\section{Pion electromagnetic form factor}
\label{sec:2}
\begin{figure}
     \includegraphics[height=5.1cm]{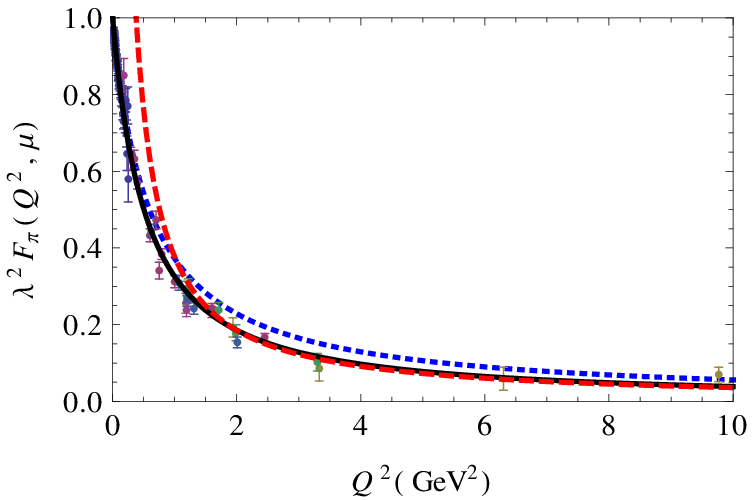} 
     \includegraphics[height=5cm]{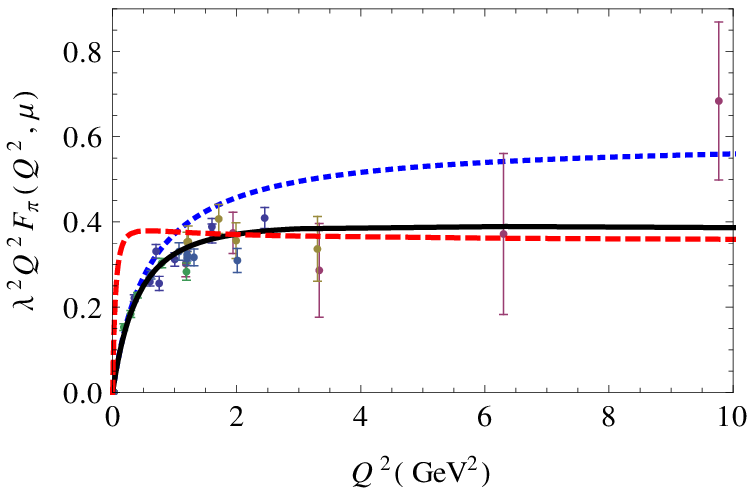} 
        \caption{The pion form factor compared to the data~\cite{Huber:2008id}. 
      The left panel shows the form factor $F_\pi (Q^2,\mu=0.14)$ when scaled with $\lambda^2=(0.42/0.14)^2$ (red dashed line) to fit the form factor $F_\pi (Q^2,\mu=0.42)$ (black solid line) together with the $\rho$-pole contribution (blue dotted line). The right panel shows the same form factors but scaled with $Q^2$. } 
\label{fig:2}
\end{figure}

As a first application of the formalism developed in~\cite{Biernat:2013fka} we use the mass function~(\ref{eq:1}) in the chiral limit for the computation of the pion electromagnetic form factor in relativistic impulse approximation (RIA)~\cite{RIA}. For simplicity, we adopt an approximated pion vertex function that is an off-shell extension of the solution in the chiral limit, together with a dressed quark current that satisfies the Ward-Takahashi identity~\cite{Gro87,Gro96,Biernat:2013aka}. Remarkably, the pion form factor $F_\pi(Q^2,\mu)$ result is insensitive to the particular choice of the quark form factors $h(p^2)$, but it depends on the pion mass $\mu$, in particular at small $Q^2$. For sufficiently large $\mu \sim m_\chi$ one expects the RIA, i.e., keeping only the spectator quark pole contribution of the triangle diagram, to be a good 
approximation for the full triangle diagram of the pion form factor, not only at high but also at low $Q^2$~\cite{Biernat:2013aka}. In particular, the value $\mu$=0.42 GeV gives the best fit to the experimental data~\cite{Huber:2008id} over the full range of $Q^2$. At large $Q^2$ we find an interesting scaling behavior between form factors of different pion masses:
\begin{eqnarray}
   F_\pi(Q^2,\lambda \mu)\stackrel{Q^2\gg\mu^2}{\simeq}\lambda^{2} F_\pi(Q^2,\mu)\,, \label{eq:scaling}
\end{eqnarray} 
 where $\lambda$ is a scaling parameter. In Figure~\ref{fig:2} the large-$Q^2$ tail of the form factor for $\mu=0.14$ GeV is scaled to fit the one for $\mu=0.42$ GeV.

It should be emphasized that this is a very simple model for the pion form factor that does not include explicit contributions from the $\rho$ meson which should be important according to vector meson dominance. Nevertheless, our result exhibits the correct monopole behavior at large $Q^2$ and is in good agreement with the data. A possible explanation would be that part of these contributions are already effectively contained in the strong quark form factors, but this interesting hypothesis needs to be further investigated.  

To summarize, the purpose of this work is to show that our model is able to give sensible results for both, the quark and the pion structure at the same time. It is clear that for a more quantitative study of the light meson properties the solution of the complete four-channel CST equation is needed, together with a more refined dressed quark current including the $\rho$ pole contribution. This will be the subject of our future program. 

\begin{acknowledgements}
This work received financial support from Funda\c c\~ao para a Ci\^encia e a Tecnologia (FCT) under grant Nos.~PTDC/FIS/113940/2009, CFTP-FCT (PEst-OE/FIS/U/0777/2013) and POCTI/ISFL/2/275. This work was also partially supported by the European Union under the HadronPhysics3 Grant No. 283286, and by Jefferson Science Associates, LLC under U.S. DOE Contract No. DE-AC05-06OR23177.
\end{acknowledgements}


\end{document}